\documentclass[12pt,preprint]{aastex}
\usepackage{epsfig}

\begin{document}

\newcommand{\lsim}   {\mathrel{\mathop{\kern 0pt \rlap
  {\raise.2ex\hbox{$<$}}}
  \lower.9ex\hbox{\kern-.190em $\sim$}}}
\newcommand{\gsim}   {\mathrel{\mathop{\kern 0pt \rlap
  {\raise.2ex\hbox{$>$}}}
\lower.9ex\hbox{\kern-.190em $\sim$}}}
\def\be{\begin{equation}}
\def\ee{\end{equation}}
\def\ba{\begin{eqnarray}}
\def\ea{\end{eqnarray}}
\def\eps{{\varepsilon}}
\def\ap{\approx}
\def\bb{\leftarrow}

%%%%%%%%%%%%%%%%%%%%%%%%%%%%%%%%%%%%%%%%%%%%%%%%%%%%%%%%%%%%%%%%%%%%%%%%%%%%%%%
\title{
{\bf Anti-GZK effect in Ultra High Energy Cosmic Rays\\ 
diffusive propagation}}

\author{R. Aloisio and V.S. Berezinsky}
\affil{INFN - Laboratori Nazionali del Gran Sasso, I--67010
Assergi (AQ), Italy}

%\email{aloisio@lngs.infn.it, berezinsky@lngs.infn.it}

\begin{abstract}
We discuss the antiGZK effect in the diffusive propagation of ultra high
energy protons in intergalactic magnetic fields,
which consists in a jump-like increase of the maximum distance 
from which ultra high energy protons can reach an observer. The position of 
this jump, $E_j \approx 2\times 10^{18}$~eV, is determined exclusively by
energy losses (transition from adiabatic to pair-production energy losses)
and it is independent of the diffusion parameters. The diffuse spectrum
presents a low-energy steepening approximately at this energy, which 
is very close to the position of the second knee observed in
the cosmic ray spectrum. The dip, seen in the universal spectrum as
a signature of the interaction with the cosmic microwave background radiation, 
is also present in the case of diffusive propagation in magnetic fields. 
\end{abstract}
%%%%%%%%%%%%%%%%%%%%%%%%%%%%%%%%%%%%%%%%%%%%%%%%%%%%%%%%%%%%%%%%%%%%%%%%%%%%%%%
\section{Introduction}
\label{introduction}
The GZK cutoff (Greisen (1966), Zatsepin and Kuzmin (1966)) is a
steepening of the ultra high energy (UHE) protons spectrum due to the 
interaction with the cosmic microwave background (CMB) radiation. The presence 
of an intergalactic magnetic field can modify the GZK cutoff up to its
absence in the case of very strong magnetic fields, Sigl et al (2004), 
Yoshiguchi et al (2003), (for a physical explanation of this effect see 
Aloisio and Berezinsky (2004)). The proton propagation in magnetic field 
can affect the observed UHE proton spectrum also at energies (much) lower 
than the GZK cutoff. The crucial parameter which determines the modification 
of the spectrum is the distance $d$ between sources. If this distance is 
much less than all propagation distances, such as energy-attenuation length, 
$l_{\rm att}$, and diffusion length $l_{\rm diff}$, the spectrum is not 
distorted and has a universal (standard) shape (Aloisio and Berezinsky (2004)).
This statement has the status of a theorem.  

All these effects depend strongly on the strength of the large-scale 
intergalactic magnetic field (IMF), the knowledge of which still 
remains poor. The modes of the UHE-proton propagation  
vary between quasi-rectilinear propagation in a weak field and
diffusive propagation in a strong magnetic field. The experimental data on
IMF and the models of origin of these fields do not
allow at present to choose even between the two extreme propagation regimes
mentioned above.

The most reliable observations of the intergalactic magnetic field are
based on the Faraday rotation of the polarized radio emission (for
reviews see Kronberg (1994), Vall\'e (1997), Carilli and
Taylor (2002)). The upper limit on the Faraday rotation measure
(RM) in the extragalactic magnetic field, obtained from the
observations of distant quasars, gives ${\rm RM} <
5~{\rm rad/m}^2$. It implies an upper limit on the extragalactic
magnetic field on each assumed scale of coherence length (Kronberg (1994),
Vall\`e (1997), Ryu et al. (1998)). For example, according to 
Blasi et al. (1999a), for an inhomogeneous universe
$B_{l_c} < 4$~nG on a scale of coherence $l_c= 50$~Mpc.

According to observations of the Faraday rotations the
extragalactic magnetic field is strongest, or order of $1~\mu$G,
in clusters of galaxies and radiolobes of radiogalaxies
(Vall\'e (1997), Kronberg (1994), Carilli and Taylor (2002)). The
largest scale in both structures reaches $l_c \sim 1$~ Mpc. Most
probably various structures of the universe differ dramatically by
magnetic fields, with very weak field in voids and much stronger
in the filaments (Ryu et al. (1998)). Superclusters seem to be too
young for the regular magnetic field to be formed in these structures on a
large scale $l_c \sim 10$~Mpc.

In the case of a hierarchical magnetic field structures in the universe,
UHE protons with $E> 4\times 10^{19}$~eV can propagate in a
quasi-rectilinear regime. Scattering of UHE protons occurs mostly
in galaxy clusters, radiolobes and filaments. Deflections of UHE
protons can be large for some directions and small for the others.
The universe looks like a leaky, worm-holed box, and correlation
with the sources can be observable (see Tinyakov and Tkachev (2001),
where correlations of ultra high energy cosmic rays (UHECR) with BLLacs 
are found). Such a picture has been suggested by Berezinsky et al. (2002b).

A promising theoretical tool to predict the IMF in large scale
structures is given by magneto-hydrodynamic (MHD) simulations.
The main uncertainty in these simulations is related to the
assumptions concerning the seed magnetic field.

The MHD simulations of Sigl et al. (2004) and Sigl et al. (2003)
favor a hierarchical structure with strong magnetic fields.
Assuming an inhomogeneous seed magnetic field generated by cosmic
shocks through the Biermann battery mechanism, the authors obtain
$\sim 100$~nG magnetic field in filaments and $\sim 1$~nG in
voids. In some cases they consider IMF up to a few micro Gauss as
allowed. In these simulations UHECR are characterized by large deflection 
angles, of the order of $20^{\circ}$, at energies up to $E\sim 10^{20}$~eV 
(Sigl et al. (2003), Sigl et al. (2004)). Thus, the scenario that 
emerges in these simulations seems to exclude the possibility of an UHECR 
astronomy. These simulations have some ambiguity related
to the choice of magnetic field at the position of the observer
(Sigl et al. (2003), Sigl et al. (2004)). The authors consider two
cases: a strong local magnetic field $B\sim 100$~nG and a weak
field $B \ll 100$~nG. The different assumptions about the local
magnetic field strongly affects the conclusions about UHECR
spectrum and anisotropy.

The essential step forward in MHD simulations has been made
by Dolag et al. (2003). In this work the Local Universe is simulated
with the observed density and velocity field. This
eliminates the ambiguity for the local magnetic field, that 
is found to be weak. The seed magnetic field, used in
this simulation, is normalized by the observed magnetic field in
rich clusters of galaxies. The results of these constrained
simulations indicate a weak magnetic fields in the universe of
the order of $0.1$~nG in typical filaments and of $0.01$~nG in
voids. The strong large-scale magnetic field, $B\sim 10^{3}$~nG,
exists in clusters of galaxies, which, however, occupy
insignificant volume of the universe. The picture that emerges
from the simulations of Dolag at el. (2003) favors a hierarchical
magnetic field structure characterized by weak magnetic fields.
UHE protons with $E> 4\times 10^{19}$~eV can propagate in a
quasi-rectilinear regime, with the expected deflection angles being very
small $\le 1^{\circ}$. 

The case  of strong magnetic fields up to $1~\mu$G has been studied in 
Sigl et al. (1999), Lemoine et al.
(1999), Stanev (2000), Harari et al. (2002), Yoshiguchi et al.
(2003), Deligny et al. (2003). The interesting features found in
these calculations are small-scale clustering of UHE particles as 
observed by Hayashida et al. (1996), Hayashida et al. (1999), 
Uchiori et al. (2000), Glushkov and Pravdin (2001), 
and absence of the GZK cutoff in the 
diffusive propagation, when the magnetic field is very strong. Many 
aspects of the diffusion of UHECR have been studied in numerical simulation 
by Casse et al. (2002).

The small-scale clustering allows to estimate the space density of the  
sources (Dubovsky et al (2000) and Fodor and Katz (2000)). The recent
Monte Carlo simulations (Yoshiguchi et al (2003), Blasi and De Marco (2004)
and Kachelrie{\ss} and Semikoz (2004)) favor a number density of the
sources $n_s \sim (1 - 3)\times 10^{-5}$~Mpc$^{-3}$ with rather large 
uncertainties (Blasi and De Marco (2004)). 

Diffusive propagation of extragalactic UHECR has been studied
already in earlier work. The stationary diffusion from Virgo
cluster was considered by Wdowczyk and Wolfendale (1979),
Giller et al. (1980) and non-stationary
diffusion from a nearby source was studied by Berezinsky et al.
(1990a), Blasi and Olinto (1999b) using the Syrovatsky solution 
(Syrovatskii (1959)) of the diffusion equation. In this case the GZK 
cutoff may be absent. 

A very interesting phenomenon, caused by propagation of UHE protons
in the extragalactic magnetic fields, has been recently found 
by Lemoine (2004). It consists in a low-energy steepening of the
spectrum of UHE 
protons at energies below $1\times 10^{18}$~eV produced by a large diffusive 
propagation time (exceeding the age of the universe) to the nearby sources. 
In this paper, we shall discuss the anti-GZK effect in diffusive
propagation of UHE protons which is responsible for this low-energy 
steepening and discuss the transition from galactic to extragalactic
cosmic rays. In our calculations we shall follow, like Lemoine (2004),   
the theoretical approach of Aloisio and Berezinsky (2004). 
%%%%%%%%%%%%%%%%%%%%%%%%%%%%%%%%%%%%%%%%%%%%%%%%%%%%%%%%%%%%%%%%%%%%%%
\section{Diffusive propagation in the analytic approach}
\label{Syr}
The analysis below
is based on the analytical solution of the diffusion equation, found
by Syrovatskii (1959). Using a distribution of sources on a lattice, the
diffuse flux can be calculated as the sum over the fluxes from the
discrete sources $i$:
\be 
\label{J-diff}
J_p^{diff}(E)=\frac{c}{4\pi}
\frac{L_p K(\gamma_g)}{b(E)} \sum_i \int_{E}^{E_g^{\rm max}}
dE_g q_{gen}(E_g) \frac{exp\left [-\frac{r_i^2}{4\lambda(E,E_g)} \right ]}
{\left ( 4\pi\lambda(E,E_g)\right )^{3/2}},
\ee
where $b(E)=dE/dt$ is the proton energy loss, summation goes over all 
lattice vertexes, $L_p$ is the proton luminosity of a source, 
$q(E_g)=E^{-\gamma_g}$ is the generation function, and $K(\gamma_g)$
is the normalization coefficient equal to $\gamma_g -2$ if 
$\gamma_g >2$ and $1/\ln(E_{\rm max}/E_{\rm min})$, if $\gamma_g=2$ 
(all energies are measured in GeV), and
\be 
\label{lambda}
\lambda(E,E_g) = \int_{E}^{E_g} d\epsilon
\frac{D(\epsilon)}{b(\epsilon)}
\ee
is the Syrovatsky variable,
which has the physical meaning of the squared distance traversed by a
proton in the observer direction, while its energy diminishes from
$E_g$ to $E$. From Eq.~(\ref{J-diff}) one can see that the
sources at distances $r > 2\sqrt{\lambda(E,E_g)}$ give negligible 
contribution to the flux.

In our calculations we shall use also the second Syrovatsky variable,
which can be understood as the time needed by a proton to diminish its 
energy from $E_g$ to $E$:
\be
\tau(E,E_g)=\int_{E}^{E_g} \frac{d\epsilon}{b(\epsilon)} ~.
\label{tau}
\ee

The Syrovatsky solution formally includes all propagation times
up to $t \to \infty$ and the generation energies are restricted 
from above only by the maximum acceleration energy 
$E^{\rm acc}_{\rm max}$ that a 
source can provide.
In our case the propagation time from a source at fixed distance $r$ 
must be smaller than the age of the universe $t_0$, and due to this condition
one more upper limit on the maximum generation energy $E_g^{\rm max}$
emerges.
This limit is given by the condition $\tau(E,E_g) \leq t_0$ and results in 
$E_g^{\rm max}(E) \leq E_g(E,t_0)$, which can be calculated also by
evolving energy backward in time from $E$ at $t=0$ to $E_g$ at $t=t_0$.

\begin{figure}[t!]
\begin{center}
\includegraphics[width=0.7\textwidth]{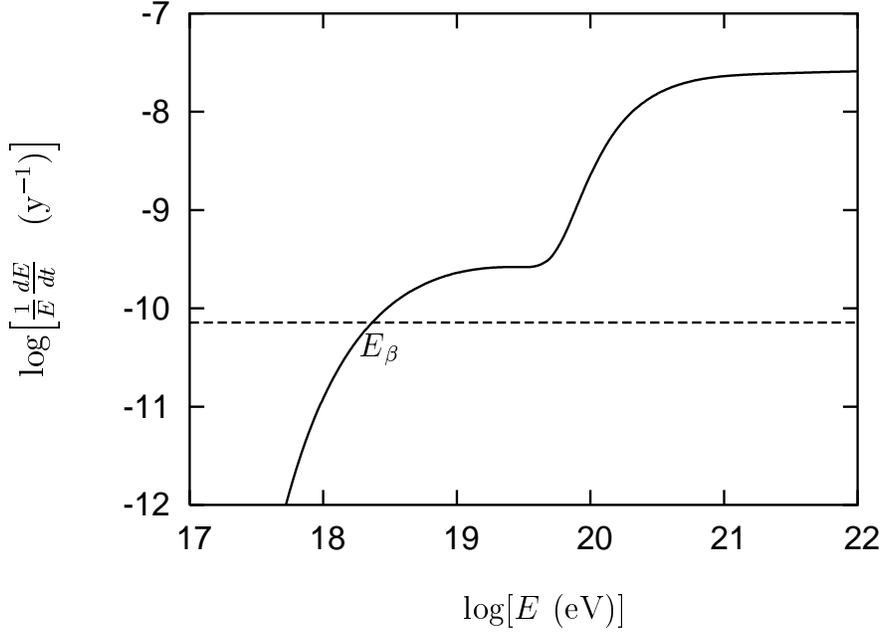}
\caption{Proton energy losses. The continuous line represents the sum of 
pair production and photopion production energy losses, while the dashed
line gives adiabatic energy losses $(1/E)(dE/dt)=H_0$. The label 
$E_{\beta}$ shows the energy where pair-production and adiabatic energy 
losses are equal.}
\label{fig1}
\end{center}
\end{figure}

The upper limit $E_g^{\rm max}$ in Eq.~(\ref{J-diff}) is then the
minimum between the two quantities: $E_g(E,t_0)$ and the maximal acceleration 
energy $E^{\rm acc}_{\rm max}$, 
\be
E_g^{\rm max}(E)={\rm min}[E_g(E, t_0),~E^{\rm acc}_{\rm max}]~.
\label{Emax} 
\ee
At small energies $E \leq 2\times 10^{18}$~eV~~  
$E_g(E, t_0) < E^{\rm acc}_{\rm max}$, while at larger energies 
$E_g(E, t_0) > E^{\rm acc}_{\rm max}$.  In the calculations below we
will assume $ E^{\rm acc}_{\rm max}=1\times 10^{22}$~eV.\\*[1mm] 
\noindent
The crucial quantity in the following discussion, the proton energy loss
$\beta(E)=(1/E)dE/dt$, is shown in Fig. \ref{fig1}. Note the
characteristic energy $E_{\beta} \approx 2\times 10^{18}$~eV, where the
pair-production energy losses $\beta_{e^+e^-}(E)$   
reach the adiabatic energy losses. 

Using these energy losses we can calculate $E_g^{\rm max}(E)$. The
results are presented in Fig. 2. At low energies $E_g(E,t_0)$ increases
due to adiabatic energy losses. At the end of this stage the increase
becomes more sharp because at large time $t$ the pair-production
energy-losses set in. Finally at $E \sim E_{\beta}$~~ $E_g(E,t_0)$ 
abruptly increases up to $E^{\rm acc}_{\rm max}$ practically by a jump. 
The jump factor is roughly given by $\exp(t_0/\bar{\tau})$, where 
$\bar{\tau}$ is the energy-loss time which diminishes as the energy 
rises with the backward time. This behavior of $E_g^{\rm max}(E)$  
is responsible for the antiGZK effect, which will be discussed in the
next Section. 

We shall specify now the diffusion coefficient $D(E)$, which
determines $\lambda(E,E_g)$ in Eq. (\ref{lambda}). In the
following discussion we shall also use the diffusion length definition as:
$l_d(E)=3D(E)/c$.  

We assume diffusion in a random magnetic field with a strength
$B_0$ on the maximum coherent length $l_c$, denoting this magnetic
configuration by $(B_0,~l_c)$.  
This assumption determines
the diffusion coefficient $D(E)$ at the highest energies when the proton Larmor
radius, $r_L(E) \gg l_c$:

\begin{figure}[t!]
\begin{center}
\includegraphics[width=0.7\textwidth]{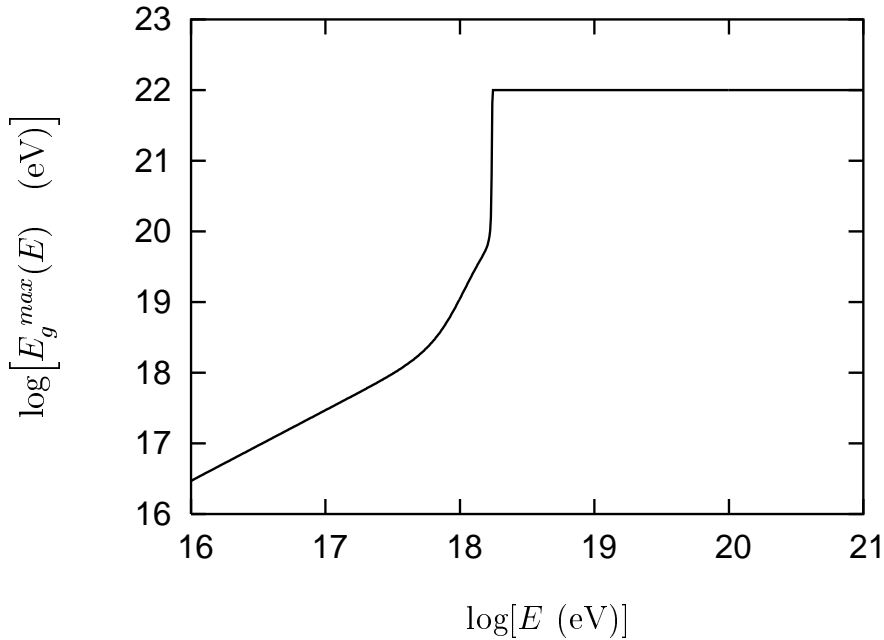}
\caption{Maximum generation energy $E_g^{\rm {max}}$ defined as 
${\rm min}[E_g(E, t_0),~E^{\rm acc}_{\rm max}]$, where 
$E^{\rm acc}_{\rm max}$ is the maximal
acceleration energy and $t_0$ is the age of the universe (see text).}
\label{fig2}
\end{center}
\end{figure}

\be
D(E) = \frac{1}{3} \frac{c r^2_L(E)}{l_c}
\label{D_HE}
\ee
At ``low'' energies, when $r_L(E) \lsim l_c$ we shall consider
three cases:\\*[1mm]
\noindent
(i) The Kolmogorov diffusion coefficient
\be
\label{D_K}
D_K(E)= \frac{1}{3}cl_c\left (\frac{r_L(E)}{l_c}\right )^{1/3},
\ee 
(ii) The Bohm diffusion coefficient
\be
\label{D_B}
D_B(E) = \frac{1}{3} c r_L(E),
\ee
(iii) An arbitrary case $D(E) \propto E^{\alpha}$, with $\alpha=2$
for the extreme energy regime.

In all cases we normalize the diffusion coefficient by
$(1/3)cl_c$ at $r_L=l_c$.
The characteristic energy $E_c$ of the transition between the 
high energy and low energy regimes is determined by the condition
$r_L(E)=l_c$ and is
\be
\label{E_c}
E_c = 0.93\times 10^{18} \left (\frac{B_0}{1~{\rm nG}} \right)
\left (\frac{l_c}{Mpc} \right)~ {\rm eV.}
\ee

The smooth transition between
the low-energy and high-energy diffusion regimes
is provided with the help of an interpolation formula for the 
diffusion length:
\be
\label{l_d}
l_{\rm diff}(E)=\Lambda_d + \frac{r_L^2(E)}{l_c}
\ee
with   
$\Lambda_d=r_L(E)$ for the Bohm diffusion and 
$\Lambda_d=l_c (r_L/l_c)^{1/3}$ for the Kolmogorov regime.

For completeness we shall give also the numerical expression for
the Larmor radius:
\be
\label{r_L}
r_L(E)= 1.08 \times 10^2 \frac{E}{1\times 10^{20}~{\rm eV}}
\frac{1~{\rm nG}}{B}~ {\rm Mpc}.
\ee
At distances $r \leq l_{\rm diff}(E)$,
the fluxes from individual sources $i$ are calculated in the 
rectilinear approximation, and the diffuse flux is given by
\be
\label{J-rect}
J_p^{\rm rect}(E)=\frac{L_p K(\gamma_g)}{(4\pi)^2} \sum_i
\frac{q_{gen}(E_g(E,r_i))}{r_i^2} \frac{d E_g(E,r)}{d E}
\ee
where $dE_g/dE$ is given 
in Berezinsky et al. (2002a). 

%%%%%%%%%%%%%%%%%%%%%%%%%%%%%%%%%%%%%%%%%%%%%%%%%%%%%%%%%%%%%%%%%%%%%%
\section{Anti-GZK cutoff}
\label{antiGZK}

In this Section we shall demonstrate that, in contrast to the GZK cutoff,
increasing of the proton energy losses at energy $E \geq 1\times 10^{18}$~eV 
results, in the case of diffusive propagation, in an {\em increase} of the 
maximal distance from which protons can arrive.  

We shall calculate below $\lambda(E, E_g^{\rm max})$, which 
according to Eq.~(\ref{lambda}) gives $r_{\rm max}^2/4$, 
where $r_{\rm max}(E)$ 
is the maximal distance from which protons with the observed energy E can 
arrive, as it follows from Eq.~(\ref{J-diff}): 
\be 
\label{r-max}
\lambda(E,E_g^{\rm max}) = \int_{E}^{E_g^{\rm max}} d\epsilon
\frac{D(\epsilon)}{b(\epsilon)}.
\ee

\begin{figure}[t!]
\begin{center}
\begin{tabular}{ll}
\includegraphics[width=0.45\textwidth]{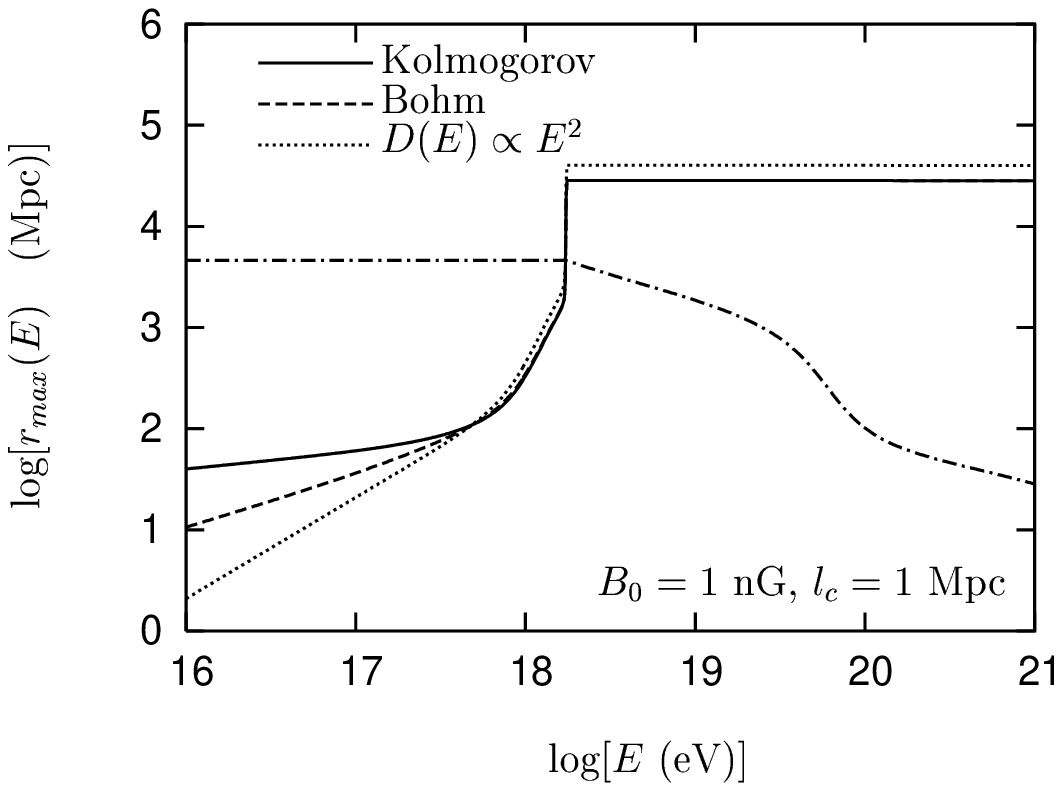}
&
\includegraphics[width=0.45\textwidth]{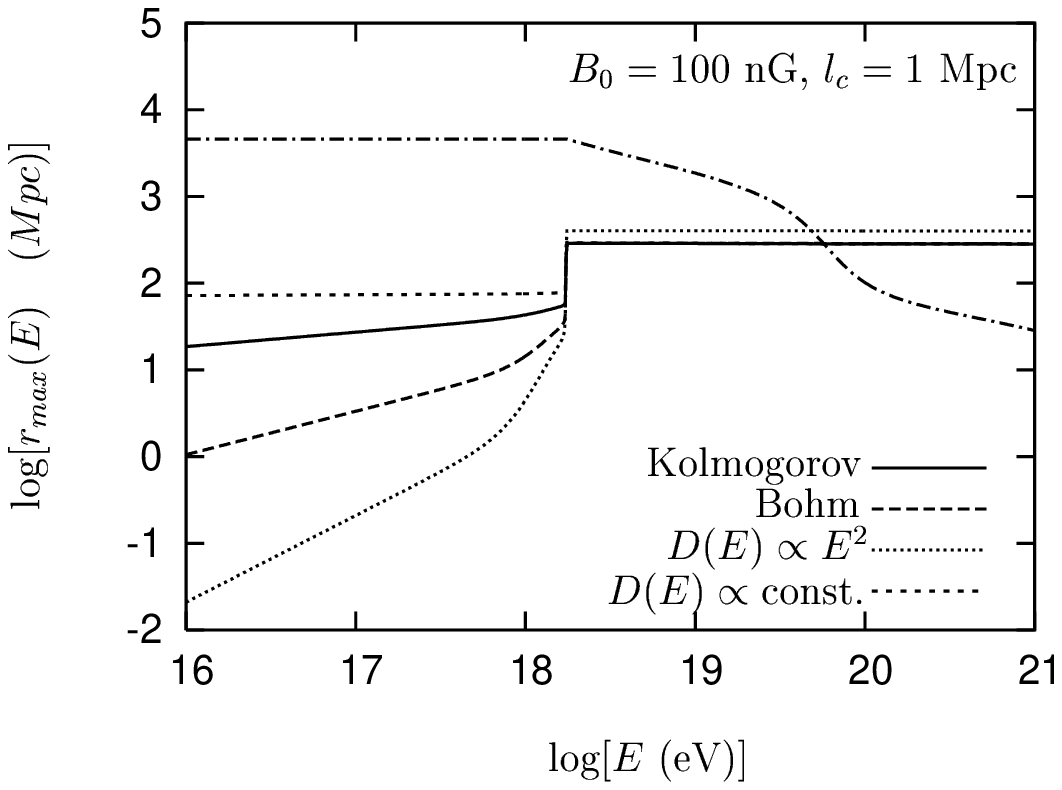}
\\
\end{tabular}
\caption{Maximal distance $r_{\rm max}(E)$  to the contributing sources
as function of the observed energy $E$. Three merging curves in the 
left-low corner give $r^{\rm diff}_{\rm max}(E)$ and the dash-dotted
curve gives $r^{\rm rect}_{\rm max}(E)$, which numerically is very
close to the energy-attenuation length $l_{\rm att}(E)=[(1/cE)dE/dt)]^{-1}$.
Continuous line is for the Kolmogorov diffusion $D(E) \propto E^{1/3}$ at 
$E\leq E_c$, dashed line - the Bohm diffusion $D(E) \propto E$, and dotted line
- the extreme high-energy diffusion $D(E) \propto E^2$. In the right
panel $D(E)=const$ case is also shown.   
Two different configurations of the magnetic field are considered: 
$B_0=1$ nG, $l_c=1$ Mpc (left panel) and $B_0=100$ nG, $l_c=1$ Mpc 
(right panel).}
\label{fig3}
\end{center}
\end{figure}
In two extreme limits, at low energies and high energies, 
$\lambda(E,E_g^{\rm max})$ can be calculated analytically.

Let us start from the low-energy case $E\ll E_{\beta}$, 
when only adiabatic energy loss operates. Using $D(E) \propto E^{\alpha}$
we obtain from Eq.~(\ref{r-max})
\be
\lambda(E,E_g^{\rm max})= \frac{D(E)}{\alpha H_0}\left [\left ( \frac
{E_g^{\rm max}}{E} \right )^{\alpha}-1 \right ].
\ee
$E_g^{\rm max}$ found from condition $\tau(E,E_g^{\rm max})=t_0$ is
$E_g^{\rm max}=E \exp(H_0 t_0)$, which results in 
\be 
r^{\rm diff}_{\rm max}(E)= 2\left ( \frac{D(E)}{\alpha H_0} \right)^{1/2}
\left (e^{\alpha H_0 t_0}-1 \right )^{1/2},
\label{Rmax-low}
\ee
where according to the WMAP data
(Spergel et al (2003)) $H_0t_0 \approx 1$.

In the extreme high-energy regime $E \geq 3\times 10^{20}$~eV~
$\tau_{\pi}^{\infty}=E(dE/dt)^{-1} \approx 4.1\times 10^7$~yr does not
depend on energy and from Eq.~(\ref{r-max}) we have
\be
r^{\rm diff}_{\rm max}(E)= \sqrt{2D_0 \tau_{\pi}^{\infty}} 
\left ( \frac{E^{\rm acc}_{\rm max}}{E_c} \right ).
\label{r-diff}
\ee
Consider now the intermediate energies, when $E$ approaches 
$1\times 10^{18}$~eV, 
but $E_g(E,t_0)$ remains less then $E_{\pi} \approx 4\times 10^{19}$~eV, where
photopion production starts. One obtains in the case $D(E) \propto E^{\alpha}$ 
\be
r^{\rm diff}_{\rm max}(E) \propto \sqrt{ D_0 \tau_{ee}}
\left ( \frac{E_g(E,t_0)}{E} \right )^{\alpha /2 },
\ee
where $\tau_{ee} \sim \beta_{e^+e^-}^{-1}$. 
In this case $r^{\rm diff}_{\rm max}(E)$ grows fast with $E$ due to 
the fast growth of $E_g(E,t_0)$ (see Fig. 2).  

When $E$ approaches $E_{\beta} \approx 2\times 10^{18}$~eV, the value 
of $r^{\rm diff}_{\rm max}$ is determined by the energy interval between 
$E_c$ and $E^{\rm acc}_{\rm max}$, where $D(E) \propto E^2$. $E_g^{max}$ there
grows by a jump to $E^{\rm acc}_{\rm max}$, and $r_{\rm max}^{\rm diff}$ also 
grows by a jump to the high energy asymptotic value given by 
Eq.~(\ref{r-diff}).

The accurate numerical calculations are displayed in Fig. \ref{fig3} for
two different magnetic field configurations (1~nG, 1~Mpc) and (100~nG, 1~Mpc),
respectively.

In a diffusive regime of propagation there is an additional upper
limit for a distance to a source, which
we shall refer to as the rectilinear maximal distance 
$r_{\rm max}^{\rm rect}(E)$.  It is defined as 
\be
r_{\rm max}^{\rm rect}(E)=\left \{ \begin{array}{ll} 
c\tau(E,E_g^{\rm max}) & {\rm if}~~ \tau < t_0, \\
ct_0                   & {\rm if}~~ \tau>t_0. \end{array}
\right.
\label{r-rect}
\ee
At small $E$ ~~ $r_{\rm max}^{\rm rect}(E)\equiv   
c\tau(E,E_g^{\rm max})=(c/H_0)\ln (E_g^{\rm max}/E)$
is larger than $ct_0$ and $r_{\rm max}^{\rm diff}(E)$, as one can see 
from Fig.~3. At large $E$~~ $r_{\rm max}^{\rm rect}(E)$ is smaller than 
$r^{\rm diff}_{\rm max}(E)$, and thus the rectilinear upper limit becomes
restrictive.  

The Syrovatsky solution (\ref{J-diff}) does not include automatically the 
restriction due to $ r_{\rm max}^{\rm rect}(E)$, because propagation time 
there varies from 0 to $\infty$. The restriction (\ref{r-rect}) must be
imposed in Eq.~(\ref{J-diff}) additionally. This restriction is valid also
in the case without magnetic field and numerically it is very close to the 
attenuation length $l_{\rm att}(E)=E (dE/dl)^{-1}$, which describes
the ordinary GZK cutoff.

Fig. \ref{fig3} illustrates the antiGZK effect which we discuss here. 
While the energy-attenuation length $l_{\rm att}(E)=E (dE/dl)^{-1}$ 
(or maximal rectilinear distance $r^{\rm rect}_{\rm max}$) diminishes with
energy $E$ and has the sharp GZK steepening at $E \sim 5\times 10^{19}$~eV, 
the diffusive maximum distance 
$r_{\rm max}^{\rm diff}(E)$ increases with energy and has a sharp 
jump at energy
$E_j \approx 2\times 10^{18}$~eV. As we discussed above, this energy is
determined entirely by energy losses and it does not depend on the 
diffusion parameters. 

The growth of $r_{\rm max}^{\rm diff}(E)$ depends on the diffusive
regime, as it directly follows from Eq.~(\ref{Rmax-low}). 

%%%%%%%%%%%%%%%%%%%%%%%%%%%%%%%%%%%%%%%%%%%%%%%%%%%%%%%%%%%%%%%%%%%%%%%%%%%%%%
\section{Results and discussion}
\label{results}
The maximum distance $r_{\rm max}(E)$ determines the number of sources
which in principle can contribute to the observed diffuse flux 
$J_p(E)$: the flux from the sources at distances $r$ larger than 
$r_{\rm max}$ is suppressed as $\exp(-r^2/r_{\rm max}^2)$.  But
inside the sphere with radius $r_{\rm max}$ the fluxes from the sources 
are suppressed by $\lambda(E,E_g)$, which is less than
$\lambda(E,E_g^{\rm max})$  
and by $E_g^{-\gamma_g}(E)$. By this reason, the 
jump in $r_{\rm max}$ does not produce a jump in the flux  
at energy $E_j$. The situation is different at 
$E < 2\times 10^{18}$~eV, where $r_{\rm max}(E)$ suppresses the diffuse
flux, restricting the number of contributing sources.

\begin{figure}[t!]
\begin{center}
\includegraphics[width=0.7\textwidth]{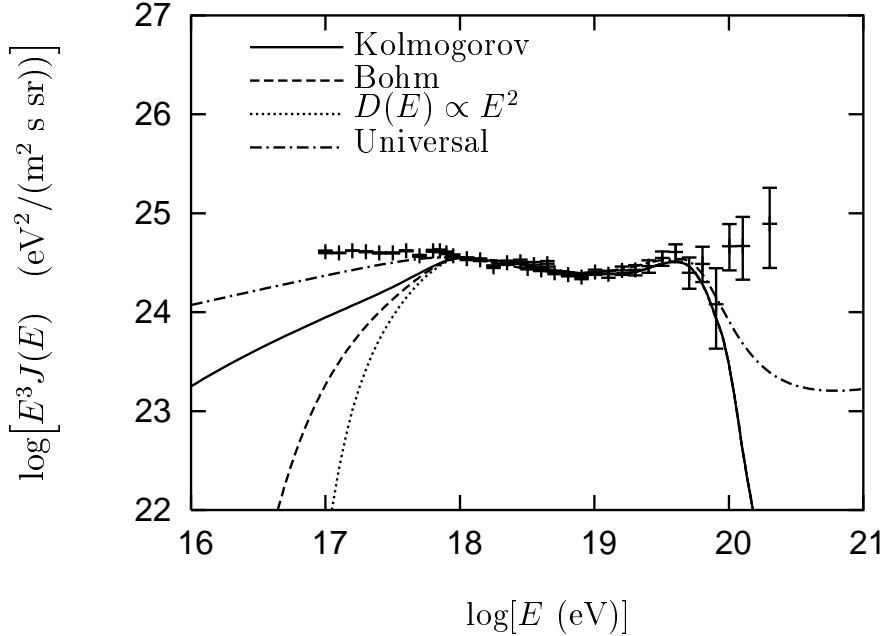}
\caption{Energy spectrum in the case of $B_0=1$ nG, $l_c=1$ Mpc and for
the diffusion regimes: Kolmogorov (continuous line), Bohm (dashed line) and
$D(E)\propto E^2$ (dotted line). The separation between sources is $d=50$ Mpc 
and the injection spectrum index is $\gamma_g=2.7$ (see text). 
The AGASA-Akeno experimental data with the universal spectrum (dash-dotted 
line) are also reported.}
\label{fig4}
\end{center}
\end{figure}

\begin{figure}[t!]
\begin{center}
\includegraphics[width=0.7\textwidth]{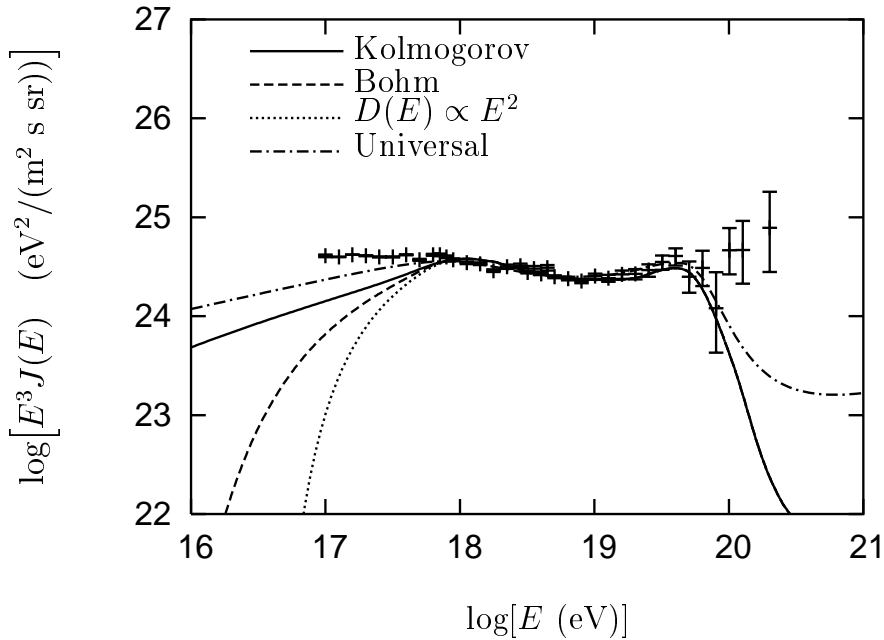}
\caption{The same as in Fig.~\ref{fig4}, but for smaller separation 
between sources $d=30$~ Mpc. Notice the better conversion to the universal
spectrum than in the case with $d=50$~Mpc, shown in Fig.~\ref{fig4}.}
\label{fig5}
\end{center}
\end{figure}

\begin{figure}[t!]
\begin{center}
\includegraphics[width=0.7\textwidth]{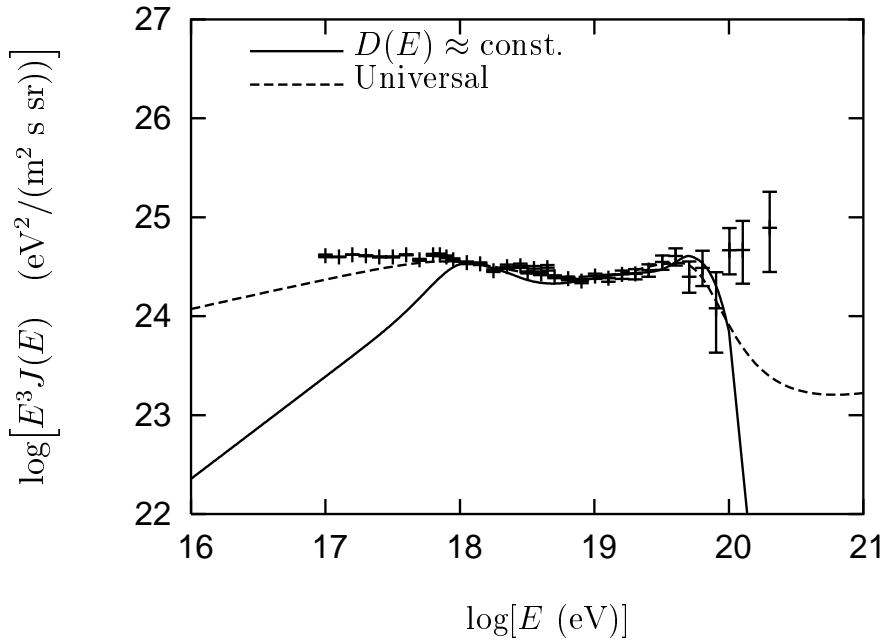}
\caption{Differential spectrum in the case of  $B_0=100$ nG, $l_c=1$ Mpc 
and for
the diffusion regime with $D(E) \simeq {\rm const}$. The separation between 
sources is $d=100$ Mpc and the injection spectrum is the standard one 
($\gamma_g=2$).
The AGASA-Akeno experimental data with the universal spectrum (dashed line) 
are also shown. The sharp cutoff in the energy spectrum at the highest
energies is due to large distances to the nearby sources 
$r \sim d \sim 100$~Mpc. 
}
\label{fig6}
\end{center}
\end{figure}

In Figs \ref{fig4},\ref{fig5},\ref{fig6} we present the calculated diffuse 
spectra 
using Eqs.~(\ref{J-diff}) and (\ref{J-rect}), in the case of 
two configurations $(B_0, l_c)$ and for different distances $d$ between 
sources.  

In our calculations the sources are located in the vertexes of a
lattice, and summation is performed within the volume limited by 
$r_{\rm max}(E)$ as described in Section 3. In fact, only the
rectilinear limit is introduced by hand, while $r^{\rm diff}_{\rm max}(E)$ at
lower energies appears automatically.

As was expected, the energy of the low-energy steepening $E_s$ is nearly   
the same for all magnetic configurations and approximately coincides with the 
cross-over of adiabatic and pair-production energy losses $E_{\beta}$, 
and with the 
position of jump $E_j$. In accordance with 
$r_{\rm max}(E)$ given by Eq.~(\ref{Rmax-low}), the flux below the 
low-energy  cutoff is the largest for the Kolmogorov diffusion (or D=const 
regime) and the 
lowest for $D(E) \propto E^2$ diffusion, with the Bohm diffusion between them. 

In the calculations for a reasonable magnetic field configuration
with $B_0=1$~nG and $l_c=1$~Mpc, we have used a separation between sources
$d=30$ Mpc and $d=50$ Mpc, which corresponds to a source space density
$3.7\times 10^{-5}$~Mpc$^{-3}$ and $8.0\times 10^{-6}$~Mpc$^{-3}$, 
respectively. As was discussed in 
the Introduction, the small-angle clustering favors a density 
$n_s \sim (1 - 3)\times 10^{-5}$~Mpc$^{-3}$ with some uncertainties. 
In the case of strong magnetic field $B_0=100$~nG we have used a larger
separation $d=100$~Mpc to improve the agreement with observations.

In Figs \ref{fig4} and \ref{fig5} we show the spectra in the case
$B_0=1$~nG and $l_c=1$~Mpc. The critical 
energy where the diffusion changes its regime is 
$E_c \sim 1\times 10^{18}$~eV, 
and the diffusion length at this energy is $l_{\rm diff} \approx 1$~Mpc.  
The best fit to the
observations is obtained for $\gamma_g=2.7$~. The energy of the steepening  
in both cases is $E_s \sim 1\times 10^{18}$~eV.  
The source luminosities $L_p$, needed to provide the observed flux are
very high, if one assumes a power-law generation spectrum from 
$E_{\rm min} \sim 1$~GeV up to $E^{\rm acc}_{\rm max}=1\times 10^{22}$~eV.  
For $d=50$~Mpc $L_p= 1.5\times 10^{49}$~erg/s and for $d=30$~Mpc~~ 
$L_p= 3.0\times 10^{48}$~erg/s. To reduce these luminosities 
one can assume that the acceleration mechanism operates starting from 
some larger $E_{\rm min}$. Then the required luminosity is reduced 
by a factor $E_{\rm GeV}^{-(\gamma_g-2)}$, which is $1.3\times 10^{-5}$
for $E_{\rm min}=1\times 10^8$~GeV, and $2.5\times 10^{-6}$ for  
$E_{\rm min}=1\times 10^9$~GeV. Another possible assumption is 
the standard spectrum $\propto 1/E^2$ at $E < E_{\rm min}$ as 
Berezinsky et al (2002b) have assumed. 

Figs \ref{fig4} and \ref{fig5} show that the dip seen in the universal
spectrum as a signature of the interaction with CMB (Berezinsky et al 2002a)
survives in the case of propagation in magnetic field with configuration
(1~nG, 1~Mpc). As will be shown below the same is true for weaker and
stronger magnetic fields.

The case of a strong magnetic field $(B_0,~l_c)=(100~{\rm nG}, 1~{\rm Mpc})$
is shown in Fig. \ref{fig6}. This is a very attractive case: the good 
agreement with the data is reached using the standard generation spectrum 
$\propto 1/E^2$ and $d=100$~Mpc. The required luminosity is  reasonable, 
$L_p=3\times 10^{45}$~erg/s for $E_{\rm min} \sim 1$~GeV and 
$E_{\rm max}=1\times 10^{22}$~eV. The diffusion coefficient used in this 
case is $D \approx const$ at $E \lsim E_c$ (the best fit in Fig. \ref{fig6}
is obtained for $D(E) \propto E^{0.02}$).  
Unfortunately, the required magnetic field is much higher than that 
obtained in the MHD simulations by Dolag et al. (2004) and 
Sigl et al. (2004), though it does not contradict the existing 
observational upper limits. 

Let us now come over to the case of very weak magnetic field 
$B_0 \sim 0.1$~nG, favored by MHD simulations by Dolag et al (2004).  
In this case $E_c \approx 1\times 10^{17} (l_c/{\rm 1~Mpc})$~eV and 
$l_{\rm diff}(E)\approx 100 E_{18}^2 ({\rm 1~Mpc}/l_c)$~Mpc. 
Therefore, for $l_c \lsim 1$~Mpc and $E \gsim 3\times 10^{18}$~eV 
the protons propagate quasi-rectilinearly in the universe. In this case
the distance between sources $d$ is less than the propagation lengths 
$l_{\rm diff}(E)$ and $l_{\rm att}(E)$, and the spectrum at least at
energies $(1-40)\times 10^{18}$~eV must be universal. 

A note of warning should be made about the validity of the Syrovatsky solution 
at $E < 1\times 10^{19}$~eV. This solution is expected to work not perfectly
well at these energies, because  it is valid only in the case when the 
energy losses $b(E)$ and diffusion coefficient $D(E)$ are
time-independent \footnote{One should be careful with inserting ad hoc
time-dependent quantities in the Syrovatsky solution (\ref{J-diff}),  
because in this case it ceases to be a solution of the corresponding
diffusion equation. For example, it is forbidden to introduce the
cosmological scaling factor $a(t)$, because it results in 
time dependent energy losses $b(E,t)$, or considering $\lambda$ in 
Eq.~(\ref{J-diff}) as function of $E, E_g$ and $t$.}.
For the above-mentioned energies this is not the case, because during 
the time of propagation the temperature of the CMB radiation changes 
appreciably, and hence the energy losses too\footnote{In our previous paper, 
Aloisio and Berezinsky (2004), we deliberately limited ourselves to 
higher energies.}. The 
diffusion equation itself should be also modified as $t \to t_0$ by the 
cosmological relations between time and distance. However, the approximate 
agreement, which we obtained (to be discussed somewhere else) between the 
Syrovatsky solution in quasi-rectilinear regime and the exact rectilinear 
propagation demonstrates the approximate validity of this solution at the 
discussed energies. 

Another argument in favor of the Syrovatsky solution as a reasonable
approximation at the discussed energies $E \lsim 1\times 10^{19}$~eV
is the convergence to the universal spectrum (compare Fig.~\ref{fig4} and 
Fig.~\ref{fig5} for $d=50$~Mpc and $d=30$~Mpc, respectively).
The universal spectrum is calculated in the case of time-dependent CMB 
temperature and for an expanding universe. The Syrovatsky solution
converges to this spectrum with accuracy better then $15\%$ 
when $d \to  3 - 5$~Mpc (to be discussed somewhere else). 

Following the papers by Berezinsky et al (2004) and Lemoine (2004),
we shall now discuss shortly the transition from galactic to
extragalactic cosmic rays. The remarkable feature of the
diffusive spectra is the low-energy steepening at the fixed energy 
$E_s \sim 1\times 10^{18}$~eV, which provide the transition from
extragalactic to galactic CR. This energy coincides approximately with the
position of the second knee $E_{\rm sk}$ and gives a non-trivial explanation
of its value as $E_{\rm sk} \sim E_{\beta}$. 

\begin{figure}[t!]
\begin{center}
\includegraphics[width=0.7\textwidth]{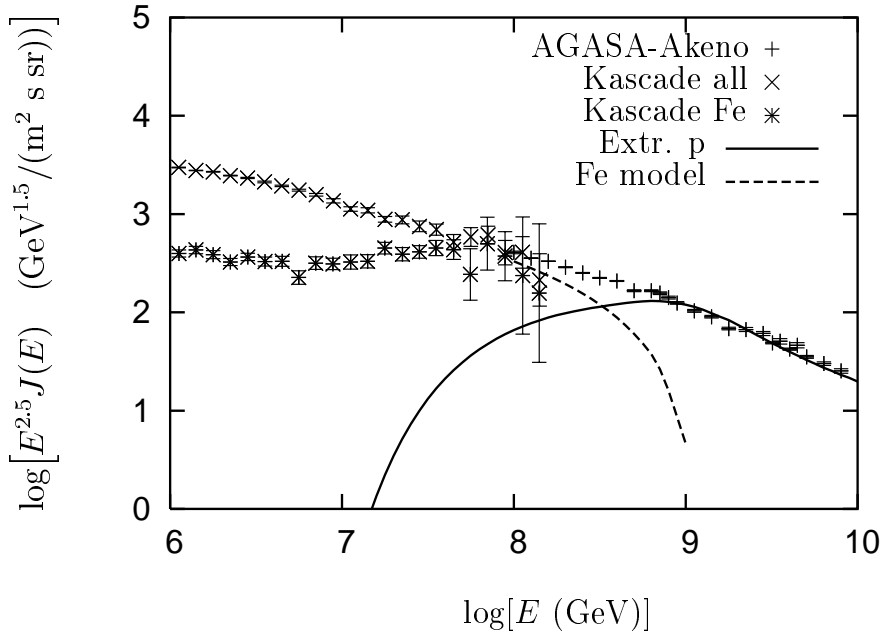}
\caption{The galactic iron-nuclei spectrum computed by subtracting 
the extralactic proton spectrum from the Akeno-AGASA data. The
extragalactic proton spectrum is taken for the case $B_0=1$ nG, 
$l_c=1$ Mpc, $d=30$ Mpc, $\gamma_g=2.7$ with the Bohm diffusion
at $E < E_c$.} 
\label{fig7}
\end{center}
\end{figure}

Like in the above-mentioned works we shall 
assume that at $E \gsim 1\times 10^{17}$~eV the galactic spectrum is
dominated by iron nuclei and calculate their flux by subtracting  
the calculated flux of extragalactic protons from all-particle Akeno 
spectrum. For these calculations we shall fix the spectrum with
magnetic configuration (1~nG, 1~Mpc), the Bohm diffusion 
at $E< E_c$ and a separation between sources on the lattice $d=30$~Mpc 
(see Fig. \ref{fig5}). The calculated spectrum of galactic iron
is shown in Fig. \ref{fig7} by the dashed curve. The fraction of 
iron-nuclei in the total flux is shown in Table 1 as a function of
energy. This prediction should be taken with caution because 
of the model-dependent calculations (assumption of the Bohm diffusion)
and uncertainties involved in the Syrovatsky solution. However, it is
interesting to note that the iron-nuclei spectrum in Fig.~\ref{fig7} 
practically coincides with the spectrum calculated by Berezinsky et al. (2004) 
for the model with the generation spectrum steepening. The iron-nuclei 
spectra in both cases are well described by the Hall diffusion
(Ptuskin et al. (1993)) in the galactic magnetic field at 
energies above the knee.

We shall compare now our results with those obtained by Lemoine (2004),
who also found the low-energy steepening of the spectrum due to
diffusion. Lemoine has limited his calculations to the case 
$B_0\sqrt{l_c} \sim 2\times 10^{-10}~{\rm G Mpc}^{1/2}$, while we
demonstrated that this phenomenon is valid for much wider range of 
parameters, for example our configuration (100~nG, 1~Mpc) corresponds 
to the Lemoine parameter two order of magnitude larger. We considered  
here a more realistic basic scale $l_c \sim 1$~Mpc and the various regimes
of diffusion, while Lemoine limited himself to the $D(E)\propto E^2$
regime only. 
We have also obtained the important result that 
the energy of the steepening is the same, $E_s \sim 1\times 10^{18}$~eV, 
for all diffusion regimes and distances between the sources, and that 
universality is determined almost entirely by the proton energy losses. 
We discussed the diffusive anti-GZK effect, which we consider as the most 
interesting observation of this work\footnote{Our work has been
performed independently from the paper by Lemoine (2004) and much
earlier. We discussed our results with Pasquale Blasi in September 2004. 
The delay with the publication was partly connected with our attempts to
overcome the problems of the Syrovatsky solution.}. 

%%%%%%%%%%%%%%%%%%%%%%%%%%%%%%%%%%%%%%%%%%%%%%%%%%%%%%%%%%%%%%%%%%%%%%%%%%%%%
\section{Conclusions}
We have analyzed in this paper the anti-GZK effect in the diffusive
propagation of ultra high energy protons. This effect consists in an 
increase of the maximum distance $r_{\rm max}(E)$, from which ultra high
energy protons can reach an observer, with an increasing of the energy $E$. 
This increase is terminated by a jump, which is located at energy 
$E_j \approx 2\times 10^{18}$~eV. The position of the jump is determined 
exclusively by energy losses (transition from adiabatic to pair-production 
energy losses) and it is independent of the diffusion parameters. The 
position of the jump practically coincides with the position of the 
aforementioned transition and gives approximately the position of the 
second knee observed in the cosmic ray spectrum (see below).
 
The observational consequences of the antiGZK effect is the 
low-energy ``cutoff'' of the diffuse spectrum, which is in fact
a steepening in the spectrum, as the GZK cutoff is. The steepening energy 
$E_s$ coincides approximately with the position of the jump, $E_s \sim E_j$, 
and it is also practically independent of the diffusion parameters, 
i.e. of the basic scale of magnetic field coherence $l_c$ and of the 
magnetic field $B_0$ on this scale. However, the shape of the steepening 
is determined by the diffusion regime: it is most steep in case 
$D(E)\propto E^2$ diffusion, most flat in case of the Kolmogorov diffusion, 
with the Bohm diffusion between them. 

In our calculations we have used the Syrovatsky solution to the diffusion 
equation, combined with the rectilinear propagation at the appropriate 
distances. The sources are located in the vertexes of a lattice with 
a spacing scale $d$ (the source separation). We have used mostly 
$d=30$~Mpc and  $d=50$~Mpc, which correspond to a space density of the sources 
$3.7\times 10^{-5}$~Mpc$^{-3}$ and $8.0\times 10^{-6}$~Mpc$^{-3}$, 
respectively. The observed small-angle clustering favors the density 
$n_s \sim (1 - 3)\times 10^{-5}$~Mpc$^{-3}$. The diffusion coefficient    
$D(E)$ is calculated for a random magnetic field with the basic scale
$l_c$ and the coherent magnetic field on this scale $B_0$. Using 
this approach we have calculated the diffusive spectra for various 
magnetic configurations $(B_0, l_c)$ and source separations $d$. 

\begin{table}[t!]
\begin{center}
\begin{tabular}{c|c|c|c|c|c}
\hline
$E$ (eV) & $10^{17}$ & $2\times 10^{17}$ & $5\times 10^{17}$ 
& $7\times 10^{17}$ & $10^{18}$ \\
\hline
$J_{Fe}/(J_{p}+J_{Fe})$ & $0.83$ & $0.66$ & $0.32$ & $0.17$ & $0.04$ \\
\hline
\end{tabular}
\caption{Fraction  of iron-nuclei in the total flux as function of the energy.}
\label{table1}
\end{center}
\end{table}

Physically the most reasonable case corresponds to a magnetic field 
configuration (1~nG, 1~Mpc) with a source separation $d=30$~Mpc and 
$d=50$~Mpc. The calculated spectra are shown in Figs \ref{fig4} and 
\ref{fig5} in comparison with Akeno-AGASA data. For a power-law generation 
spectrum with $\gamma_g=2.7$ the agreement is good, but needs too high 
luminosity of the sources $L_p$, if the power-law spectrum starts with 
low energy $E_{\rm min} \sim 1$~GeV. This problem can be amiliorated assuming 
higher values of $E_{\rm min}$.

The calculated diffusive spectra in the energy interval 
$(1 - 80)\times 10^{18}$~eV agree perfectly well with the universal 
spectrum and experimental data, showing the presence of the dip caused 
by $e^+e^-$ production.

An interesting case is given by the diffusion in strong magnetic field 
with basic configuration (100~nG, 1~Mpc) and source separation d=100~Mpc. 
In this case (Fig \ref{fig6}) the best fit of the spectrum is obtained for 
the standard acceleration spectrum $Q(E)\propto 1/E^2$ and 
$E_{\rm min} \sim 1$~GeV. The required luminosity is 
$L_p=3\times 10^{45}$~erg/s. Up to energy $E \sim 1\times 10^{20}$~eV the 
predicted spectrum agrees with data of both detectors, AGASA and HiRes. 
The sharp cutoff at $E \sim 1\times 10^{20}$~eV is produced due to large 
distances $r \sim d$ to the nearby sources. For the explanation of the AGASA 
excess at $E \gsim 1\times 10^{20}$~eV a new component of ultra high 
energy cosmic rays (e.g. from superheavy dark matter, see Aloisio et al 2004) 
is needed.

At energies $E < E_s$, where $l_{\rm diff}$ becomes much smaller than $d$,
the diffusive spectrum exhibits a steepening in contrast to the universal 
spectrum (see Figs \ref{fig4} - \ref{fig6}). 

The steepening of the spectrum at $E_s \sim 1\times 10^{18}$~eV 
provides a natural transition from galactic to extragalactic cosmic rays. 
This energy coincides with the second knee observed in cosmic rays spectra by 
most of the detectors. While the energy of the transition $E_s$ 
(and thus position of the second knee) is predicted in a model independent way,
the shape of the proton spectrum below $1\times 10^{18}$~eV and the fraction 
of galactic iron nuclei are model dependent: they differ for various 
diffusion regimes.
%%%%%%%%%%%%%%%%%%%%%%%%%%%%%%%%%%%%%%%%%%%%%%%%%%%%%%
\section*{Acknowledgments}
\noindent
We are grateful to Pasquale Blasi for useful discussion. We thank the 
transnational access to research infrastructures (TARI) program trough the 
LNGS TARI grant contract HPRI-CT-2001-00149.

\end{document}